# Multifunctional Spin Logic Gates In Graphene Spin Circuits

Dmitrii Khokhriakov[1], Shehrin Sayed[2,3], Anamul Md. Hoque[1], Bogdan Karpiak[1], Bing Zhao[1], Supriyo Datta[4], Saroj P. Dash[1,5*]

[1]Department of Microtechnology and Nanoscience, Chalmers University of Technology, SE-41296, Göteborg, Sweden.
[2]Department of Electrical Engineering and Computer Sciences, University of California-Berkeley, Berkeley, CA 94720, USA.
[3]Materials Science Division, Lawrence Berkeley National Laboratory, Berkeley, CA 94720, USA.
[4]Electrical and Computer Engineering, Purdue University, West Lafayette, Indiana 47907, USA.
[5]Graphene Center, Chalmers University of Technology, SE-41296, Göteborg, Sweden.

## Abstract

All-spin-based computing combining logic and nonvolatile magnetic memory is promising for emerging information technologies. However, the realization of a universal spin logic operation representing a reconfigurable building block with all-electrical spin current communication has so far remained challenging. Here, we experimentally demonstrate a reprogrammable all-electrical multifunctional spin logic gate in a nanoelectronic device architecture utilizing graphene buses for spin communication and multiplexing and nanomagnets for writing and reading information at room temperature. This gate realizes a multistate majority spin logic operation (sMAJ), which is reconfigured to achieve XNOR, (N)AND, and (N)OR Boolean operations depending on magnetization of inputs. Physics-based spin circuit model is developed to understand the underlying mechanisms of the multifunctional spin logic gate and its operations. These demonstrations provide a platform for scalable all-electric spin logic and neuromorphic computing in the all-spin domain logic-in-memory architecture.

Keywords: spin logic, spin majority logic gate, universal spin logic, nanomagnet, spin summation, logic-in-memory, graphene, spin transport

**Corresponding author**: Saroj P. Dash, Email: saroj.dash@chalmers.se

**Introduction**

Utilizing electron spin as a state variable for logic-in-memory architecture has significant potential, allowing to design circuits beyond von Neumann architecture and realize low-power, compact, and fast information technologies[1,2]. It offers a possibility to combine the nonvolatile memory with spin logic to achieve highly efficient integrated components for information technology and artificial intelligence[3]. Multiple proposals to realize spin logic devices have been put forward, including spin transistors[4], magnetologic gates[5], all-spin switches[6], magnetoelectric spin-orbit devices[7], and concepts based on the magnetic domain wall motion[8] and spin-wave interference[9,10]. Such devices promise beneficial application-specific architectures with reduced device footprint and improved energy efficiency compared to conventional charge-based technologies.

The magnetologic and all-spin logic concepts are attractive due to their all-electrical control, reconfigurability, and possible integration with the existing technologies[11,12]. Furthermore, utilizing weighted summation of spin currents based on the vector nature of spins can be useful for advanced neuromorphic computing[13]. The realization of these devices has been severely restricted by the short spin coherence length in metallic spin interconnects. However, with the advent of long-distance spin communication in large-area multiterminal graphene spin circuits[14–16], the realization of spin-based computational schemes should be feasible. Up to now, only a few basic building blocks for spin logic are experimentally realized, such as a spin "exclusive or" (XOR) operation using spin current in graphene and silicon[17,18]. However, these devices do not allow the realization of a complete set of logical operations. In contrast, a spin majority gate, where the majority of the input states determine the output logic state, was identified as a logic primitive for spintronic circuits[19]. Such a spin majority gate paired with an inversion functionality forms a universal Boolean set representing a reconfigurable building block for future information technologies[20]. However, an experimental realization of a comprehensive and reconfigurable spin majority logic gate based on pure spin current has so far been lacking.

Here, we demonstrate an all-electrical multifunctional spin logic gate based on pure spin currents operating at room temperature. We take advantage of efficient spin communication between nanomagnetic input and output terminals in a large-area graphene integrated spin circuit. The basic mechanism relies on spin multiplexing within the graphene channel, i.e., the intrinsic capability to add and cancel spin densities in a linear regime with electrical control of individual spin inputs using tunable bias currents. The multifunctional spin logic gate can be reconfigured to achieve spin majority logic (sMAJ), AND, OR, and XNOR logic operations, and their inversion (NOT), according to the information encoded in the magnetization direction of the magnetic electrodes. The results are confirmed with circuit simulations in SPICE where the device structure was modeled using physics-based spin circuit models[21,22] to gain detailed insights on the experimental observations. These findings represent a milestone in the development of spin logic gates and open ways to realize multi-terminal circuits combining both spin-based logic and magnetic memory technologies in a large-scale integrated spin circuits.

**Main**

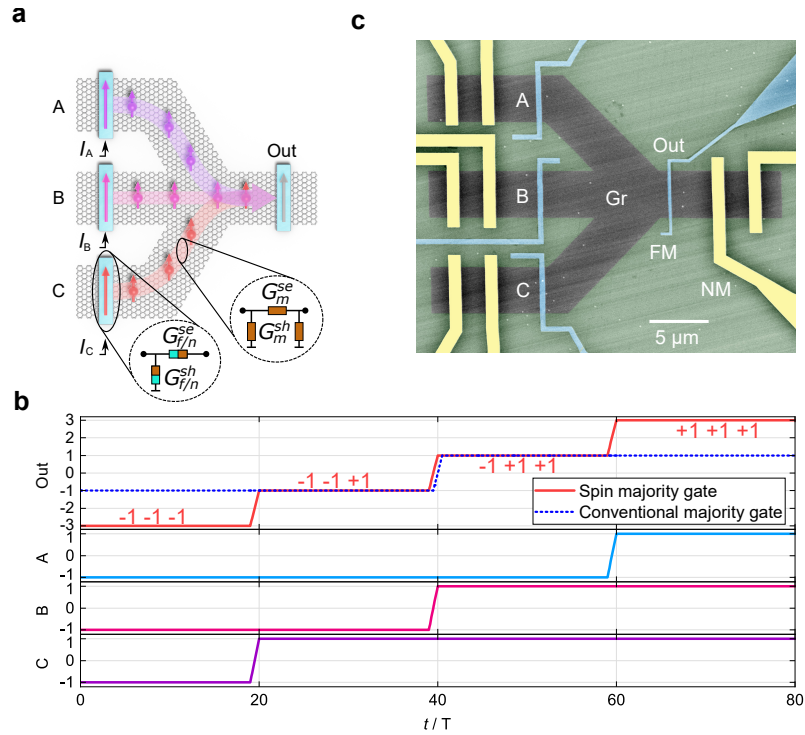

**Figure 1. Spin majority logic device in a graphene-based spin integrated circuit. a,** A schematic of the spin majority gate with a three-input fan-in circuit, where the ferromagnetic contacts are used for writing and reading the spin information, and the graphene channel is used for information communication via pure spin currents. The majority gate logic output is true (or "1") when half or more arguments are true, and false (or "0") otherwise. To simulate the device, spin circuit models employ conductance matrices $G$ to capture spin transmission through FM/Gr interfaces and spin diffusion and relaxation in graphene. **b,** A simulation comparing the output signal of a conventional charge-based majority gate and a spin majority gate for different combinations of input states. Spin-based realization shows multiple signal levels, which provide additional information on the number of inputs that constitute the majority. **c,** A colored scanning electron microscope (SEM) picture of the spin majority logic gate. The device consists of a three-input (A, B, C) and one-output (O) multi-terminal fan-in CVD graphene (Gr) circuit with ferromagnetic electrodes (Co/TiO$_2$, blue color). The type of injected spin polarization, spin-up ("1") or spin-down ("0"), can be controlled by changing the magnetization of electrodes or by applying electrical bias currents of opposite polarities (+$I$ or -$I$) to achieve spin injection or extraction. The resultant spin state is probed as an output voltage signal after spin transport and mixing in the graphene buses. The non-magnetic electrodes (Au/Ti, yellow) are used as reference contacts.

The spin majority logic gate operation (Fig. 1a) exhibits an output voltage *V* > 0 when the majority of the inputs are "1" and *V* < 0 when the majority of the inputs are "0", similar to conventional majority gates. However, the output of the spin majority logic gate is multistate due to the unique addition and cancellation of spin currents in the graphene channel. Thus, while the sign of the output voltage determines the majority spin type, the voltage level can quantify the number of inputs that constitute the majority. We have analyzed the spin majority logic gate operations using four-component (one charge and z, x, and y polarized spins) physics-based spin circuit models for the non-magnetic graphene (Gr) channel and the ferromagnetic (FM) inputs, as shown in Figure 1a. The model captures both the spin diffusion and relaxation in the graphene channel, and electron transmission and spin mixing at the FM/Gr interface (see details in the Supplementary Note 1). The simulations on a three-input spin majority gate show four-state output with two positive and two negative states, see Figure 1b. Each input magnet electrically injects a spin current $i_S$ that propagates along the channel and reaches the output magnet, which acts as a spin selective voltage probe that converts the spin current in the channel into a charge voltage. When all the inputs produce spin-down or spin-up currents ("0" or "1" states), they add up in the channel to $-3i_S$ or $+3i_S$ respectively, and the output voltage is $\propto \pm 3i_S$. When one of the magnets has an opposite logic state compared to others, a spin current cancellation leads to the output voltage of $\propto \pm i_S$. Thus, the voltage level indicates whether all three or only two of the inputs have the same logic, and the output sign reveals if the majority state is "0" or "1".

The basic building blocks for the realization of multifunctional spin logic operations are the channel materials that allow excellent spin communication in complex circuit architectures, and ferromagnetic elements that represent non-volatile memory and act as source and drain for spin-polarized carriers. Graphene is an ideal material to be utilized as a spin interconnect in the proposed spin logic concepts because its low intrinsic spin-orbit coupling allows long spin coherence lengths[23]. In particular, the recent advances with the demonstration of robust room-temperature spin interconnect functionality of the large-area chemical vapour deposited (CVD) graphene with more than 30 µm spin communication distance, and the realization of complex spin circuit architectures make it a prime candidate to build scalable spintronic technologies[14,24–26]. We take advantage of such excellent spin transport properties in CVD graphene on a 4-inch Si/SiO$_2$ wafer for the fabrication of spin logic devices. Figure 1c shows a scanning electron microscope (SEM) picture of the nanofabricated CVD graphene-based multifunctional spintronic device contacted with ferromagnetic TiO$_2$/Co (FM) and nonmagnetic Ti/Au (NM) electrodes (See fabrication details in the Methods section).

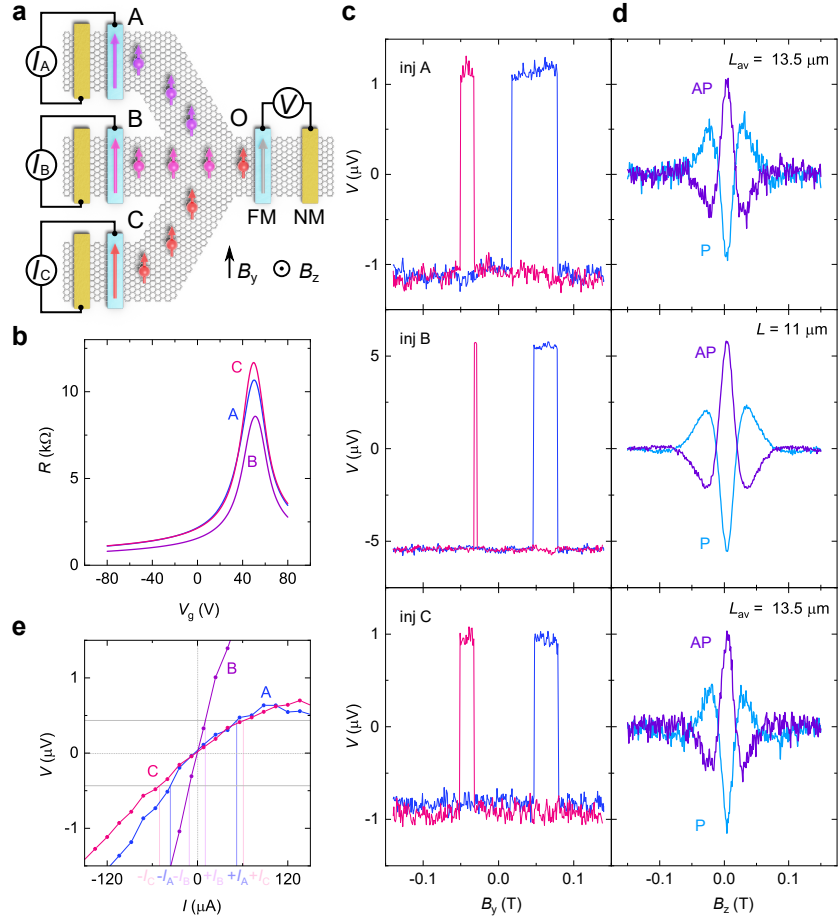

**Figure 2. Spin communication in the graphene circuit**. **a,** A schematic with a measurement geometry for characterizing the spin transport in graphene branches A, B, and C. **b,** Gate dependence of the graphene channel resistance in each branch. **c,d,** Spin valve and Hanle spin precession measurements for each channel with inputs from A, B, and C with magnetic field sweeps in $B_y$ and $B_z$ directions, respectively. The measurements are performed for parallel (P) and antiparallel (AP) magnetic orientations of input and output ferromagnetic contacts. The measurements are performed with the bias current $I$ = -200µA. **e,** A bias dependence of the spin valve signal magnitude ($V$) for each injector. The grey horizontal lines represent the selected output voltage level (limited by the nonlinearity observed for injectors A and C), and the corresponding vertical lines determine the values for positive and negative bias currents used in the majority gate measurements. All the measurements were performed at room temperature.

## Spin communication in a graphene circuit

First, we measured spin communication in the graphene circuit connected with magnetic memory elements (Figure 2a). Figure 2b shows the gate dependence of graphene channel resistance (Dirac curve), demonstrating uniform doping in all branches. The spin transport measurements were performed in the nonlocal geometry (Figure 2a), where a bias current $I$ applied between an input FM and graphene creates spin accumulation under the contact, which then diffuses through the graphene interconnect and is detected by the output magnet in the form of a voltage signal $V$. To perform spin valve measurements we sweep an in-plane magnetic field $B_y$ along the easy-axis of the magnets, which switches the magnetization direction of FMs between the parallel (P) and antiparallel (AP) configurations and results in sharp changes in the detected spin voltage, as shown in Figure 2c. A small spin-independent linear background was subtracted from the data, which can be suppressed by optimizing the device design, or using an auxiliary voltage source[17,27–29]. Next, Hanle spin precession measurements were performed by applying an out-of-plane magnetic field $B_z$, which induces spin precession and dephasing in graphene (Figure 2d). The curves are fitted with a standard Hanle equation[30] to estimate spin parameters in all channels with spin lifetime $\tau \sim 250$ ps, spin diffusion coefficient $D \sim 0.036$ m$^2$s$^{-1}$, and spin diffusion length $\lambda \sim 3$ μm. This demonstrates good spin communication properties in the multi-terminal graphene spin circuit.

The observed variations in the magnitude of spin voltage in different graphene branches can be attributed to different channel lengths, small variations in spin transport parameters, and disparity in tunnel spin polarization of each input FM electrode. To achieve the equal magnitude of spin signals from all branches, we calibrated the bias current values by measuring the spin signal magnitude as a function of bias current for each channel, as shown in Figure 2e. We observed nonlinearity in the bias dependence, which is a common feature seen in ferromagnetic tunnel contacts and can be attributed to the thermal effects, magnetic proximity, or energy-dependent spin-resolved density of states at the Gr/FM interface[31,32]. This nonlinearity in our devices results in different values for the bias currents required to achieve the same value of the spin voltage at the negative and positive sides. By choosing a desired output voltage level (marked by the gray horizontal lines in Figure 2e), we determine the positive and negative values of the bias currents $I_A$, $I_B$, and $I_C$ to be used for the following bias-controlled spin logic measurements. The output spin voltage magnitude is sufficient for the present spin logic operations, however, it can be further enhanced by downscaling the graphene spin circuit dimensions and improving the spin polarization of the ferromagnetic contacts by using hBN[33] or novel oxide[34] tunnel barriers.

## Spin Multiplexing and XNOR spin logic operation

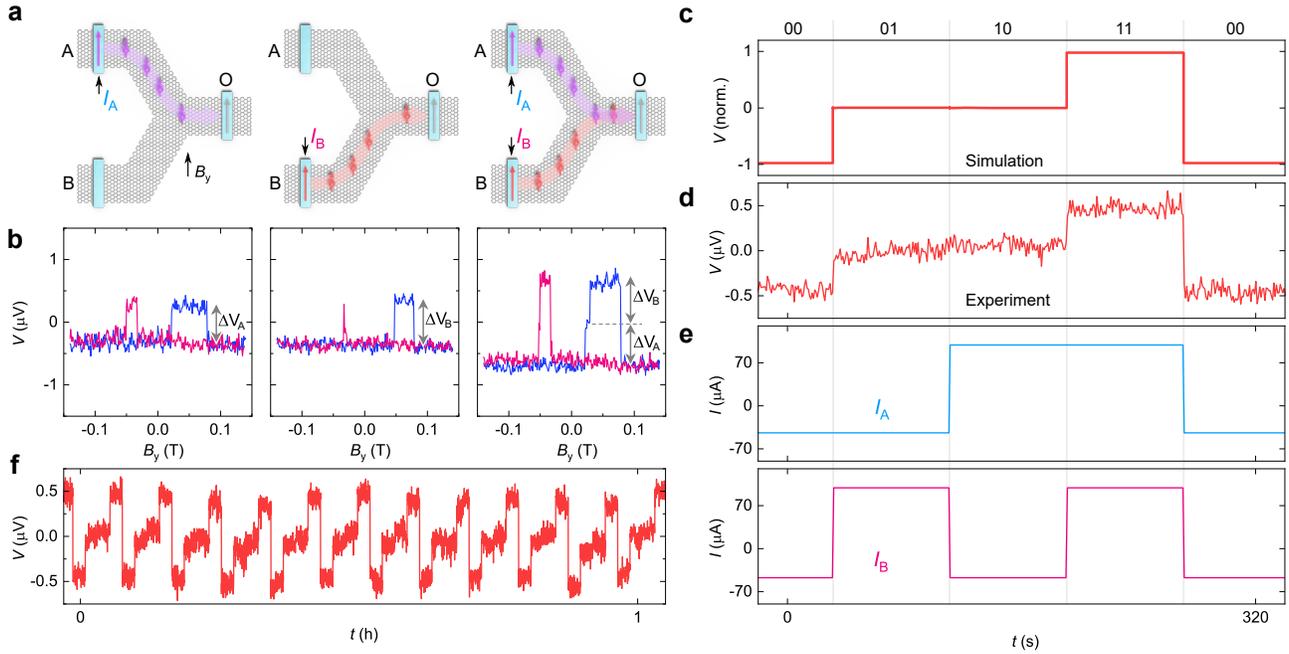

**Figure 3. Spin multiplexing and XNOR spin logic operation in a two-input fan-in device. a,b** Spin valve signals obtained using individual inputs A and B, and a spin addition or a spin cancellation operation performed by simultaneously injecting spins from both branches of the device. **c,d**, The output spin voltage at the detector as a function of time, the panels show spin circuit-based simulation and experiment respectively. **e,** The sequence of the applied bias currents to the two input contacts A and B. **f**, The device performance over several cycles. All the measurements were performed at room temperature.

Multiplexing is one of the basic spin logic operations, allowing to fan-in and fan-out spin signals and perform multiple information processing operations. As spin currents decay over the spin diffusion length, it is crucial to investigate how they combine and divide in a multi-terminal graphene spin circuit. We realized the spin addition and subtraction operations using a simultaneous injection of spin currents from two different ferromagnetic input terminals (Figure 3a). Figure 3b shows the spin valve signals obtained separately in two different branches ($\Delta V_A$ and $\Delta V_B$) of the device, and a resulting output spin valve signal with both inputs working at the same time ($\Delta V_A+\Delta V_B$). The magnitude of the output spin valve in this case represents the sum of the spin signals from both inputs, demonstrating a linear spin addition regime.

As the application of the magnetic field is not a viable option for integrated spin circuits, the polarity of injected spins is alternatively controlled by electric bias. In this method, the magnetization of all contacts remains fixed throughout the measurement, whereas the sign of the bias current determines whether the inputs work in the spin injection or spin extraction regimes. When using the bias current to control the state of each injector, we define the input state of "1" as the application of positive bias current leading to spin-up injection and the state of "0" as the application of the negative bias current leading to the spin-down injection (practically achieved by spin-up extraction). These operations result in the spin mixing and an output spin-dependent voltage as shown in Figures 3c,d, from SPICE simulation and experiment respectively for different input states defined by bias currents as a function of time (Figure 3e). The output of this two-input spin logic gate shows three stable states corresponding to the regimes where both input magnets inject spin-up currents ("11"), spin-down currents ("00"), or opposite spin currents ("01" and "10"). For the same logic states at both inputs, same type of spin currents are injected in the channel which add up to provide a high voltage state at the output. On the other hand, when the inputs inject opposite spin currents, they get canceled, resulting in a zero output voltage. This resembles the logic operation of conventional negated exclusive OR (XNOR) logic gate with output "0" and "1" logic states being zero and high voltages respectively; however, the spin XNOR gate in Fig. 3 distinguishes between "11" and "00" states according to the output voltage sign. Also, the output voltage sign can be inverted by reversing the magnetization states of all injectors or the detector before the experiment (see Fig. S2). The stability of the spin addition device operation is demonstrated in Figure 3f, where the output voltage over multiple operation cycles is shown. Such fan-in and fan-out[14] multiplex geometries show that spin current can be precisely controlled by the spin injection bias and spin resistances of each graphene branch in a circuit. These experiments demonstrate the possibility to define electrically the spin input states of a spintronic logic gate, which is a vital ingredient for developing spin-based logic circuits.

## Spin majority logic gate operation

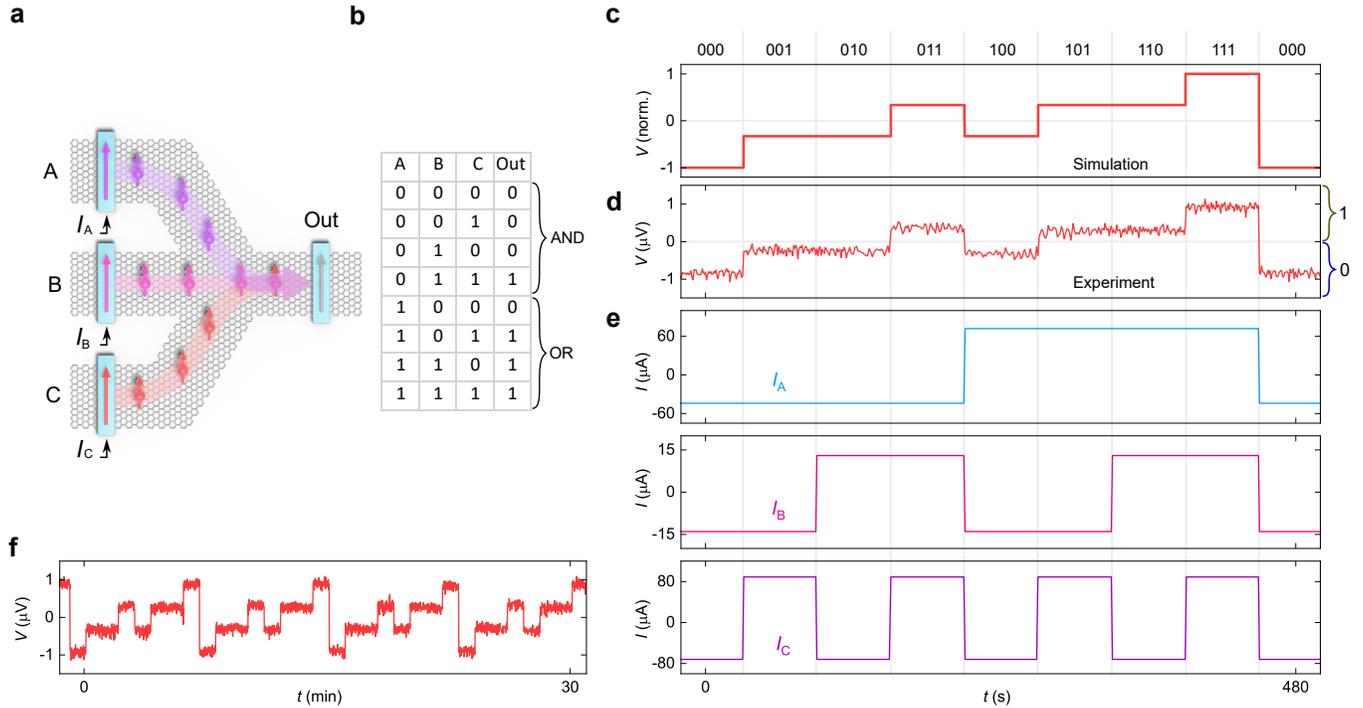

**Figure 4. Spin majority gate operation with 3-input fan-in device. a,** A schematic of the spin majority gate device, where all three spin inputs (A, B, C) are used simultaneously, and the majority of the injected spin polarizations define the nonlocal output (O) spin voltage. **b,** The truth table for the spin majority gate. The output is true when at least two of the inputs are true. By fixing any of the inputs as "0" or "1", the output for the remaining two inputs represents an AND or an OR operation, demonstrating the reconfigurability of the gate. **c,d,** A spin circuit model-based simulation and experimental realization of the spin majority gate operation achieved by applying a sequence of the pulse bias currents to all inputs ($I_A$, $I_B$, $I_C$, panel **e**) as a function of time to probe all possible logic states. The $+I$ results in spin injection and $-I$ leads to spin extraction. **f,** The device performance over several cycles. All the measurements were performed at room temperature.

Finally, after establishing the essential ingredients such as spin communication and spin multiplexing functionalities, we investigated the all-electrical spin majority gate operation using three inputs of the device simultaneously (Figure 4a). According to the truth table of the majority gate (Figure 4b), its output should be "1" if at least two of the three inputs are in the "1" state, and "0" otherwise. Thus, a majority logic gate represents digital functions on the basis of majority decisions, and can be re-programmed by fixing one of the three input variables to "0" or "1", such that it can function as an AND or OR gate for the remaining two inputs. The negative and positive bias currents were used to represent the "0" and "1" states of the inputs, creating spin currents of opposite polarization in the graphene circuit (spin-down and spin-up, respectively). The magnetization of the input and output FMs are aligned in the same direction and the bias current magnitude was chosen from the bias dependence calibration. Figure 4c shows spin circuit model-based

SPICE simulation results of the three-input spin majority gate, which provide insights into the underlying mechanisms of the experimental results presented in Figure 4d. Both simulations and experiments use a sequence of currents to the three inputs as shown in Figure 4e, which goes through all possible logic combinations. The output signal switches between four stable voltage levels, two levels corresponding to the stronger majority cases: ("111" and "000") and the other two levels corresponding to weaker majority cases: ("011", "101", "110") and ("001", "010", "100"). When charge-based background contributions are eliminated, we can add a comparator and label the output as logical "1" when the resulting nonlocal spin voltage signal is positive, and "0" otherwise. In this convention, it is clear to see that the resulting signal indeed complies with the majority gate truth table and that the gate can be re-programmed to realize AND or OR operations by fixing any of the inputs as "1" or "0" respectively when the input magnets are parallel to the output magnet. The majority gate can also be controlled by the magnetic field, however, this approach does not allow to probe all possible combinations of the input states as the magnetization switching order is determined by the individual coercive fields of the FMs (see Fig. S3). However, this approach can be used to achieve the inversion of the output signal (see Fig. S4) when the input magnets are antiparallel to the output magnet, thus allowing to realize NAND and NOR functions. The majority gate operation was stable over several cycles as shown in Figure 4f, demonstrating such graphene integrated circuits as robust and promising building blocks for spin logic applications.

**Discussion**

In the quest for energy-efficient and high-performance non-von Neumann platforms, such spin-based devices have the potential to emulate human brain-like functionalities with the collocation of memory and processing units. Furthermore, the ferromagnetic elements in the device provide nonvolatile functionality as well as the potential for reprogrammability, which makes these graphene circuits promising for logic-in-memory computation by integration with the standard magnetic memory technologies. The nonvolatility of the spin inputs and outputs substantially reduces energy consumption in computation because spin logic circuits do not consume power when idle and do not need reloading of data after power-off. The multi-input spin majority logic gate provides an attractive platform for scalable neuromorphic computing allowing weighted summation of input signals in the all-spin domain where the weights will be determined by the magnetic configuration at the inputs. Since the graphene interconnects communicate information via pure spin currents with reduced thermal dissipation, the spin-based neuromorphic architectures implemented with magnets on graphene can be more energy-efficient than a purely charge-based resistive hardware. Our realization of the multifunctional logic gate using pure diffusive spin currents is scalable using advanced lithography and fully compatible with industrial fabrication processes. In principle, this makes this concept more efficient compared to charge-based counterparts and can reduce the device footprint in application-specific circuits. The output voltage magnitude is largely determined by the device lateral dimensions and the efficiency of spin injection/detection, and can be boosted by employing highly spin-polarized ferromagnetic tunnel contacts or van der Waals heterostructures of graphene with 2D magnets[35]. Furthermore, the device performance and functionalities can be improved by incorporating spin-transfer[36] or spin-orbit torque for magnetization switching and novel topological spin-orbit materials with large charge-spin conversion efficiency[37].

## Summary

In conclusion, we demonstrate a prototype all-electrical multifunctional spin logic device operating with pure spin currents at room temperature. This was made possible by realizing spin multiplexing operation in spin circuits fabricated using large-area CVD graphene channels and demonstrating fan-in operations with spin currents. The all-electrical operations are achieved by using tunable bias currents to control individual spin inputs of XNOR and spin majority logic (sMAJ) gates, which can also be reconfigured to perform (N)AND and (N)OR operations. In order to understand the underlying mechanisms of the experimentally realized spin logic gates, we use spin circuit model based multiphysics simulations in SPICE. The simulations agree well with the experimental results and provide insights on how spin current addition and cancellation provide multistate output. These results open the door for the development of spin-based information processing technology with promising applications in next-generation logic-in-memory computing architectures and brain-inspired computing.

## Methods

**Fabrication of devices**

The large area CVD graphene (from Grolltex Inc.) was grown on Cu foil and transferred onto a 4-inch wafer with 285 nm $SiO_2$ on $n^{++}Si$ substrate, with prefabricated alignment markers. The graphene circuits were patterned by electron beam lithography (EBL) and oxygen plasma etching. Both the nonmagnetic Ti (10 nm)/Au (80 nm) and ferromagnetic ($TiO_2$/Co) tunnel contacts were defined using two EBL processes, followed by electron beam evaporation of metals and lift-off processes. The ferromagnetic contacts were produced by electron beam evaporation of ~3 Å of Ti in two steps, each followed by *in situ* oxidation in a pure oxygen atmosphere to form a $TiO_2$ tunnel barrier layer. In the same chamber, we deposited 40 nm of Co, after which the devices were finalized by lift-off in acetone at 65 °C.

**Spin and charge transport measurements**

The final devices were wire-bonded and measurements were performed in a cryostat at room temperature with a magnetic field and a sample rotation stage in vacuum conditions. In the spin majority gate measurements, individual Keithley 6221 current sources were connected to each of the three input FM (TiO$_2$/Co) electrodes to apply bias currents, and the nonlocal output voltage was detected by a Keithley 2182A nanovoltmeter; the gate voltage was applied using a Keithley 2400 source meter. The three-terminal resistance of FM contacts was in the range of 1-2kΩ and 0.3-1kΩ for NM Ti/Au contacts, whereas the graphene exhibited sheet resistance of 565 Ω. An application of the gate voltage between the n$^{++}$Si substrate and graphene across the 285 nm oxide layer of SiO$_2$ was used to quantify the carrier concentration in graphene. The spin logic measurements were performed at zero gate voltage, which corresponds to the hole doping with the carrier density of n = 3.8x10$^{12}$ cm$^{-2}$ and field-effect mobility of µ = 2900 cm$^2$/Vs.

**Simulation**

The SPICE model for the spin majority gate was constructed by dividing the structure into several sections and representing the sections with corresponding physics-based spin circuit models for non-magnetic graphene and magnetic TiO$_2$/Co contacts. The spin circuit models consist of four components: one charge and z, x, and y polarized spins. The boundary conditions for charge and spin terminals are assigned according to the experimental configurations. The charge terminals are connected to current sources for the input magnets, while the spin terminals are grounded to consider the spin absorption within the magnets. The charge terminal of the output magnet is open-circuited, and the spin terminals are grounded. The boundary conditions for spins terminals in the graphene channel are all open-circuited. For the input branches in the graphene channel, the boundary terminals for charge are grounded so that the charge current injected through the input magnets can complete their paths. For the output branch of the graphene channel, the boundary terminal for the charge current is open. We apply pulse current sources to the input magnets and observe the open circuit charge voltage at the charge terminal of the output magnet in the transient simulations done in HSPICE.

**Acknowledgements**

The authors acknowledge financial support from EU Graphene Flagship (Core 3 No. 881603), Swedish Research Council VR project grants (No. 2016–03658), 2D TECH VINNOVA competence center (No. 2019-00068), Graphene center, EI Nano, and AoA Materials program at Chalmers University of Technology. We acknowledge the help of staff at Quantum Device Physics and Nanofabrication laboratory in our department at Chalmers.

**Data availability**

The data that support the findings of this study are available from the corresponding authors on a reasonable request.

**Author information**

Dmitrii Khokhriakov[1], Shehrin Sayed[2,3], Anamul Md. Hoque[1], Bogdan Karpiak[1], Bing Zhao[1], Supriyo Datta[4], Saroj P. Dash[1,5*]

**Affiliations**

[1]Department of Microtechnology and Nanoscience, Chalmers University of Technology, SE-41296, Göteborg, Sweden.
[2]Department of Electrical Engineering and Computer Sciences, University of California-Berkeley, Berkeley, CA 94720, USA.
[3]Materials Science Division, Lawrence Berkeley National Laboratory, Berkeley, CA 94720, USA.
[4]Electrical and Computer Engineering, Purdue University, West Lafayette, Indiana 47907, USA.
[5]Graphene Center, Chalmers University of Technology, SE-41296, Göteborg, Sweden.

**Contributions**

D.K. fabricated and measured the graphene spin logic devices and circuits. A.M.H, B.K., B.Z. helped in device fabrication and optimization. D.K. and S.P.D conceived the idea and designed the experiments, analyzed and interpreted the experimental data, compiled the figures. S.S and S.D performed the spin circuit simulations. D.K., S.S. S.D., S.P.D wrote the manuscript with inputs from all co-authors. S.P.D. supervised the research project.

**Corresponding author**

Saroj P. Dash, Email: saroj.dash@chalmers.se

**Competing interests**

The authors declare no competing interests.

# Supplementary information

## Multifunctional Spin Logic Gates In Graphene Spin Circuits

Dmitrii Khokhriakov[1], Shehrin Sayed[2,3], Anamul Md. Hoque[1], Bogdan Karpiak[1], Bing Zhao[1], Supriyo Datta[4], Saroj P. Dash[1,5*]

[1]Department of Microtechnology and Nanoscience, Chalmers University of Technology, SE-41296, Göteborg, Sweden.
[2]Department of Electrical Engineering and Computer Sciences, University of California-Berkeley, Berkeley, CA 94720, USA.
[3]Materials Science Division, Lawrence Berkeley National Laboratory, Berkeley, CA 94720, USA.
[4]Electrical and Computer Engineering, Purdue University, West Lafayette, Indiana 47907, USA.
[5]Graphene Center, Chalmers University of Technology, SE-41296, Göteborg, Sweden.

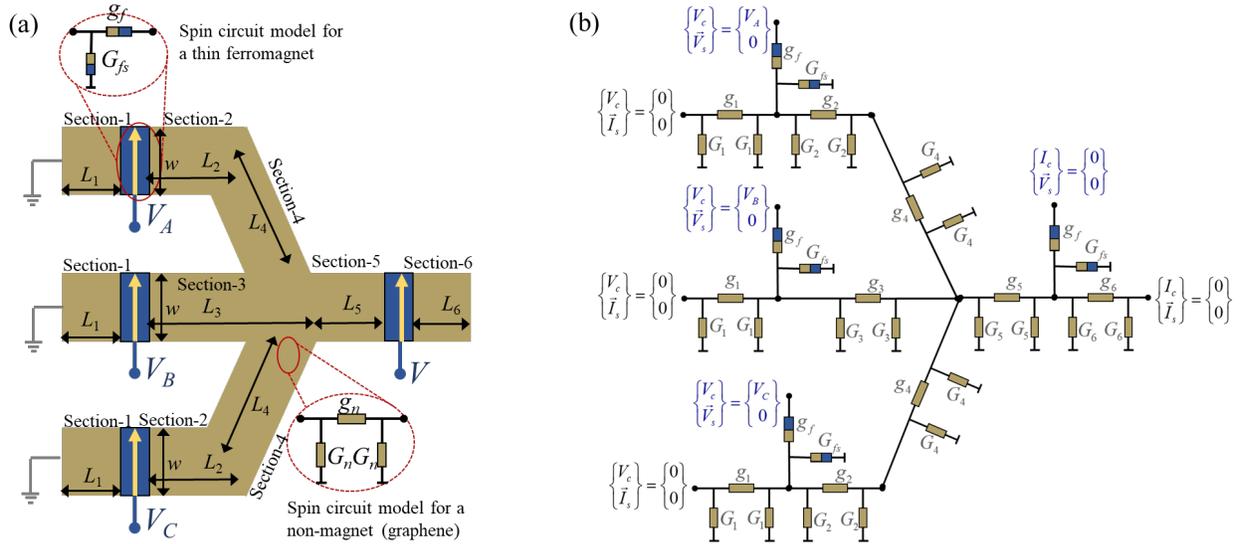

**Figure S1. Spin circuit simulations. a,** Structure of the spin majority gate and spin circuit models for the corresponding material layers. **b,** Spin-circuit modeling for the spin majority gate where the graphene channel is broken down into six sections and each section is represented by the corresponding model. The models are connected using standard circuit laws to represent the experimental structure and proper boundary conditions are applied according to the experimental conditions.

**Note 1. SPICE Model for the Spin Logic Gates.**

In this subsection, we discuss the numerical simulations of the proposed graphene-based spin majority gate. We use physics-based spin circuit models for the thin ferromagnets and the non-magnetic graphene channels [1,2]. These models are four-components: one charge and z, x, and y polarized spins and can be attached together using standard circuit laws to analyze various structures. The model for the non-magnetic material is represented by a series conductance ($g_n$) and two shunt conductances ($G_n$), which are given by

$$g_n = \frac{w}{R_{\text{sheet}} L_n} \begin{bmatrix} 1 & 0 & 0 & 0 \\ 0 & \frac{L_n}{\lambda} \text{csch} \frac{L_n}{\lambda} & 0 & 0 \\ 0 & 0 & \frac{L_n}{\lambda} \text{csch} \frac{L_n}{\lambda} & 0 \\ 0 & 0 & 0 & \frac{L_n}{\lambda} \text{csch} \frac{L_n}{\lambda} \end{bmatrix}, \quad (1)$$

$$G_n = \frac{w}{R_{\text{sheet}} L_n} \begin{bmatrix} 0 & 0 & 0 & 0 \\ 0 & \frac{L_n}{\lambda} \tanh \frac{L_n}{\lambda} & 0 & 0 \\ 0 & 0 & \frac{L_n}{\lambda} \text{csch} \frac{L_n}{\lambda} & 0 \\ 0 & 0 & 0 & \frac{L_n}{\lambda} \text{csch} \frac{L_n}{\lambda} \end{bmatrix}, \quad (2)$$

where $w$ is the width of the graphene wire, $L_n$ is the length of the $n^{\text{th}}$ section, $\lambda$ is the spin-diffusion length of the graphene which is ~ 3 µm (see main text), and $R_{\text{sheet}}$ is the sheet resistance of the graphene which is ~ 565 Ω. The series conductance takes into account the longitudinal charge and spin transport. The shunt condutances take into account the spin-flip process in the nonmagnetic channel.

For the thin ferromagnetic contacts, we use the ferromagnet-nonmagnet interface model [1,2] which is also represented by a four-component series conductance and a four-component shunt conductance, as given by

$$g_f = [\Re] \begin{bmatrix} G_0 & pG_0 & 0 & 0 \\ pG_0 & G_0 & 0 & 0 \\ 0 & 0 & 0 & 0 \\ 0 & 0 & 0 & 0 \end{bmatrix} [\Re]^{\dagger}, \quad (3)$$

$$G_{fs} = [\Re] \begin{bmatrix} 0 & 0 & 0 & 0 \\ 0 & 0 & 0 & 0 \\ 0 & 0 & 2G_r^{\uparrow\downarrow} & 2G_i^{\uparrow\downarrow} \\ 0 & 0 & -2G_i^{\uparrow\downarrow} & 2G_r^{\uparrow\downarrow} \end{bmatrix} [\Re]^\dagger, \quad (4)$$

where $G_0$ is the contact conductance of the ferromagnet, $p$ is the polarization of the ferromagnet, $G_r^{\uparrow\downarrow}$ and $G_i^{\uparrow\downarrow}$ is the real and imaginary parts of the interface spin-mixing conductance, and $[\Re]$ is the rotational matrix to take into account the magnetization vector.

We breakdown the graphene wires in the spin majority gate into six sections (see Fig. 1(a)), represent them with the four-component nonmagnetic model in Eqs. (1)-(2), which is implemented as a sub-circuit in HSPICE. We connect the charge and spin terminals of each of the sections in a modular fashion according to the standard laws of circuits, see Fig. 1(b), to construct the spin majority gate structure. Note that each node is 4-component. We represent the input and output magnets using the model for ferromagnet-nonmagnet interface in Eqs. (3)-(4). The models for three input magnets (A, B, and C) are connected at the right ends of graphene section-1 in each of the input hands. The left ends of the graphene section-1 have the boundary condition ($V_c$, $I_z$, $I_x$, $I_y$) = (0, 0, 0, 0), i.e., grounded for charge and open circuit for all spins. The boundary conditions for the input magnets model are ($V_c$, $V_z$, $V_x$, $V_y$) = ($V_A$, 0, 0, 0), ($V_B$, 0, 0, 0), and ($V_C$, 0, 0, 0), respectively. Thus, a charge current flows through the graphene section-1 in each of the input hands.

The output ferromagnet model is connected to the left end of the graphene section-6 and the boundary condition of the output ferromagnet model is ($I_c$, $V_z$, $V_x$, $V_y$) = (0, 0, 0, 0), i.e. open circuit for charge but short circuit for all-spins. The right end of the graphene section-6 has the boundary condition of ($I_c$, $I_z$, $I_x$, $I_y$) = (0, 0, 0, 0), i.e., open circuit for both charge and all-spins. Note that we apply short circuit boundary conditions for the spin terminals in all of the magnet models to take into account the spin relaxation in the magnet. Here lengths of the six graphene sections are taken as $L_1$ = 3.67 µm, $L_2$ = 1.75 µm, $L_3$ = 7.62 µm, $L_4$ = 9.33 µm, $L_5$ = 3.25 µm, and $L_6$ = 3.67 µm, calculated based on the scanning electron microscope picture of the device in Fig. 1 of the main text.

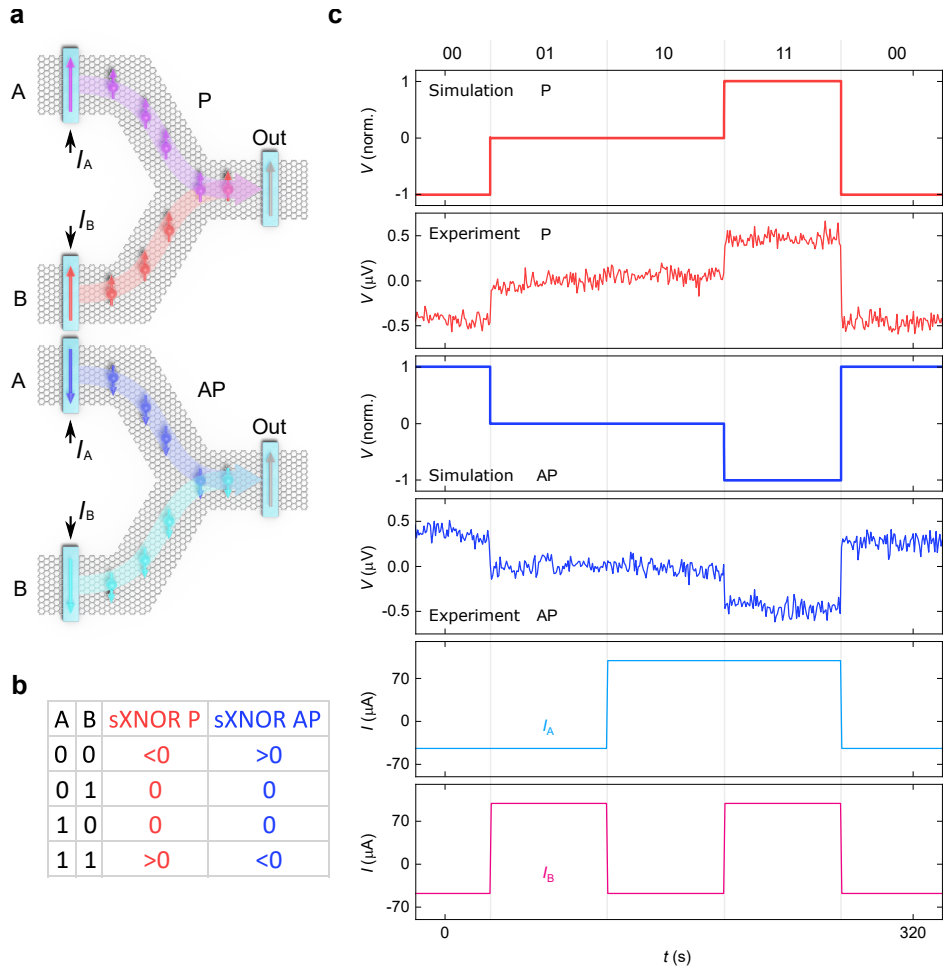

**Figure S2. Signal inversion by magnetization reversal. a,** Schematics of the spin summation device where the two injector FMs are prepared in a parallel (P) or antiparallel (AP) magnetization orientation state with respect to the output FM (Out). **b,** Truth table of the spin XNOR operation for the parallel and antiparallel orientation of the magnetization. **c**, Output spin signal of a spin summation device obtained in simulation and experiment for the P and AP configurations.

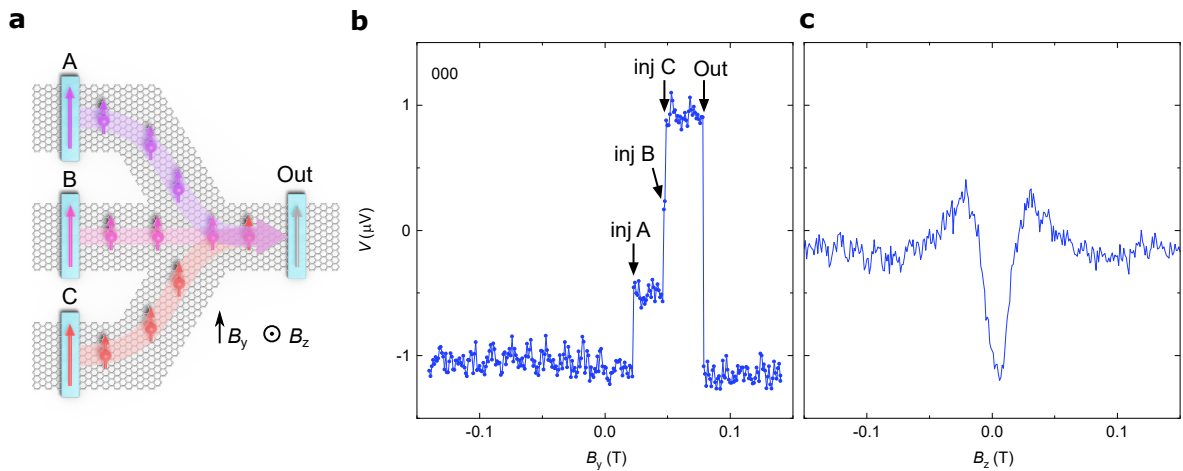

**Figure S3. Spin majority logic gate control with magnetic fields. a,** A schematic of the spin majority gate device. **b**, A spin valve measurement of the spin majority gate device obtained by sweeping the in-plane magnetic field $B_y$. The signal exhibits four different voltage levels, which were traced to the switchings of the individual FM electrodes A, B, C, and Out using additional measurements for each injector to determine their coercive fields. **c**, A Hanle spin precession curve obtained in the "000" state of the spin majority logic gate. The upper and lower injectors (A and C) are expected to produce spin precession signal contributions with a narrower width compared to the middle injector (B) because of their greater channel length, however, this difference is not strong enough to be clearly distinguished in the measurement.

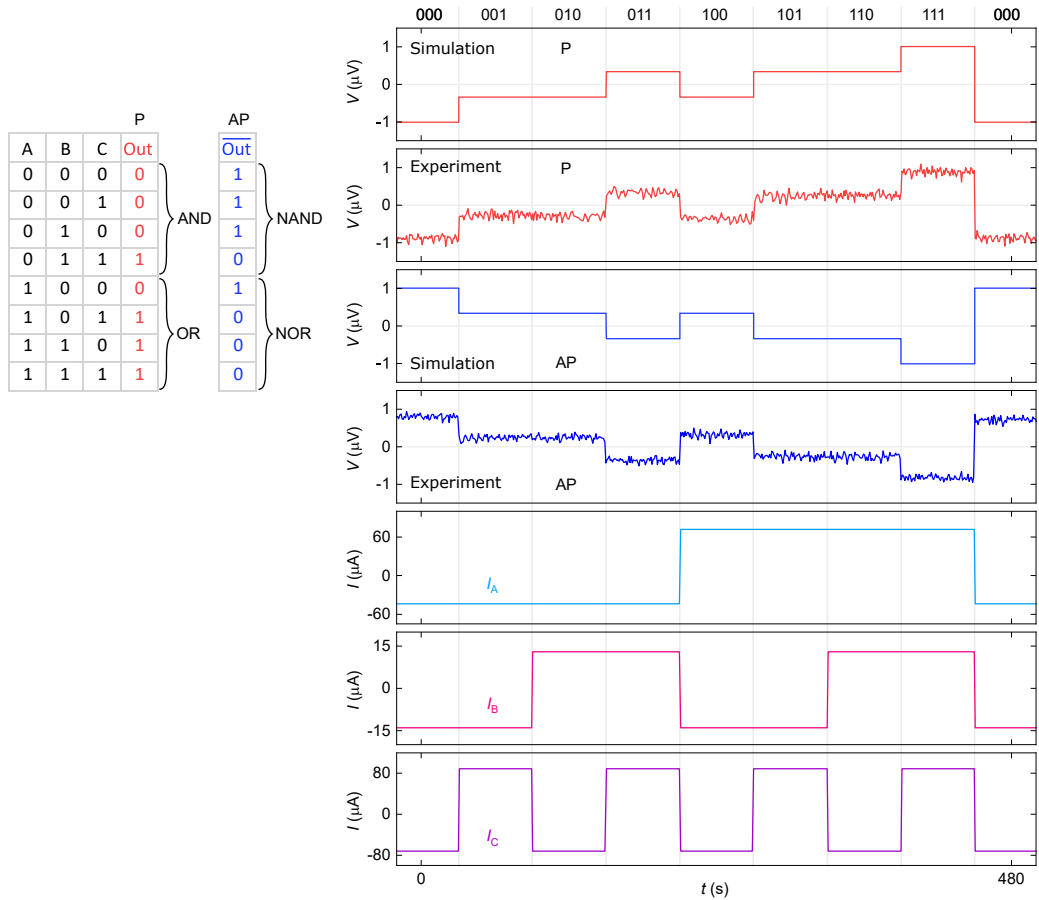

**Figure S4. Signal inversion by magnetization reversal in a majority gate.** Output spin signals obtained in simulation and experiment for a spin majority gate in the P and AP configurations, and a sequence of the bias currents applied to the three inputs. Output signal inversion in the AP configuration is achieved by reversing the magnetization orientation of all inputs prior to the experiment, allowing this gate to be reconfigurable to perform AND, OR, NAND, and NOR operations, fulfilling a complete Boolean logic set.